\documentstyle{mn}

\def\spose#1{\hbox to 0pt{#1\hss}} 
\def\simlt{\mathrel{\spose{\lower 3pt\hbox{$\mathchar"218$}}
     \raise 2.0pt\hbox{$\mathchar"13C$}}}
\def\simgt{\mathrel{\spose{\lower 3pt\hbox{$\mathchar"218$}}
     \raise 2.0pt\hbox{$\mathchar"13E$}}}

\begin{document}

\title[The supersoft source RX~J0513.9-6951]
 {Optical variability of the LMC supersoft source RX~J0513.9-6951 from 
MACHO Project photometry}

\author[C. Alcock et al.]
{C. Alcock$^{1,2}$, R.A. Allsman$^{3}$, D. Alves$^{1}$, T.S. Axelrod$^{1,4}$,
D.P. Bennett$^{1,2}$,  \and P. A. Charles$^{5}$, 
K.H. Cook$^{1,2}$, K.C. Freeman$^{4}$, K. Griest$^{2,6}$, J. Guern$^{2,6}$,
\and M.J. Lehner$^{2,6}$, M. Livio$^{7}$, 
S.L. Marshall$^{2,8}$, B.A. Peterson$^{4}$, M.R. Pratt$^{2,8}$, \and 
P.J. Quinn$^{4}$, A.W. Rodgers$^{4}$, 
K.A. Southwell$^{5}$, C.W. Stubbs$^{2,8,9}$, \and 
W. Sutherland$^{5}$ and D.L. Welch$^{10}$\\
$^1$ Lawrence Livermore National Laboratory, Livermore, CA 94550, USA\\
$^2$ Center for Particle Astrophysics, University of California, Berkeley, 
CA 94720, USA\\
$^3$ Supercomputing Facility, Australian National University, Canberra, A.C.T. 
0200, Australia\\
$^4$ Mt.  Stromlo and Siding Spring Observatories, 
Australian National University, Weston, A.C.T. 2611, Australia\\
$^5$ University of Oxford, Department of Astrophysics, Nuclear \& Astrophysics
Laboratory, Keble Road Oxford, OX1 3RH\\
$^6$ Department of Physics, University of California,
San Diego, CA 92093, USA\\
$^7$ Space Telescope Science Institute, 3700 San Martin Drive, Baltimore, 
MD~21218, USA\\
$^8$ Department of Physics, University of California, Santa Barbara, CA 93106, 
USA\\
$^9$ Departments of Astronomy and Physics, University of Washington, Seattle, 
WA 98195, USA\\
$^{10}$ Department of Physics and Astronomy, Mc~Master University, Hamilton, 
Ontario, Canada, L8S~4M1\\}

\date{Received
      in original form	}

\maketitle

\begin{abstract}
Using the exceptional monitoring capabilities of the MACHO project
we present here the optical history of the LMC supersoft source (SSS) 
RX~J0513.9-6951, for a 
continuous 3 year period. Recurring low states, in which the 
optical brightness drops by up to a magnitude, 
are observed at quasi-regular intervals. This provides a 
crucial insight into the 
nature of the SSS and, in 
particular, a chance to investigate the poorly understood behaviour of their 
accretion discs. 
Analysis of the 
high state data reveals a 
small modulation of semi-amplitude $\sim$~0.02 magnitudes at 
P$=$0.76278$\pm$0.00005~days, a period which is 
consistent with the current ``best'' suggested spectroscopic value.
\end{abstract}


\section{Introduction}
{\it ROSAT} observations have considerably enlarged the new class of high 
luminosity X-ray objects, 
the so-called ``supersoft sources'' (SSSs)  
characterised by their EUV temperatures (Tr\"{u}mper 1992). 
SSSs were 
first detected 
in the Large Magellanic Cloud in 1979-1980 with the 
{\it Einstein} X-ray Observatory (Long, Helfand \& Grabelsky, 1981). 
Until recently, little 
progress had been made in determining the exact nature of these elusive 
systems, the high level of X-ray absorption rendering them undetectable 
in the 
Galactic plane. Most 
currently known SSS are therefore extragalactic and, as such, they are 
optically faint.  

The current inventory of supersoft objects (Hasinger 1994; 
Cowley et al.\ 1996; Kahabka \& Tr\"umper 1996) 
is of the order of 11 in the Magellanic 
Clouds, 15 in M31 and 7 Galactic sources, with candidates existing also in 
M101, NGC253 and M33. 
It appears now, through ROSAT observations, 
that the SSS do not form a strictly homogeneous class, some 
having been identified with a planetary nebula nucleus (Wang 1991), a 
PG~1159 star (Cowley et al.\ 1995) 
and symbiotic systems (e.\ g.\ Hasinger 1994). 
However, the bolometric luminosities 
of these objects ($L_{\rm bol} \sim~10^{37}$~erg\,s$^{-1}$) are 
typically 
an order of magnitude less than the original {\it Einstein} sources, CAL~83 
and CAL~87. 

Among the {\it ROSAT} discoveries are several systems, including 
RX~J0513.9-6951 (hereafter RX~J0513-69) 
which show the hallmarks of the original LMC SSS. Typically, 
$L_{\rm bol} \sim~10^{38}$~erg\,s$^{-1}$ and T$_{\rm bb} \sim 30$~eV. 
Spectroscopic studies (Smale et al.\ 1988; Pakull et al.\ 1988) 
readily identified them as low mass X-ray 
binaries (LMXBs), yet the nature of the compact object and the source of the 
soft emission proved elusive. 
Previously, X-rays of such low energy had only been observed in certain types 
of cataclysmic variables (CVs), but these accreting white dwarf binary systems 
are typically about a million times fainter than the SSS (see e.\ g.\ 
Warner 1995).  

In 1990, CAL~87 
was proposed to be a black hole binary, on the basis 
of a radial velocity analysis (Cowley et al.\ 1990). 
The following year, a paper appeared 
advocating the scenario of a
neutron star accretor, shrouded in a dense cocoon of ionised matter 
(Greiner, Hasinger \& Kahabka 1991).  
However, neither model seemed to provide a natural explanation for the 
extremely soft X-ray emission. 

The most significant progress occurred when van den Heuvel 
et al.\ (1992) proposed a model for the SSS which involved a 
white dwarf primary 
undergoing accretion at a rate $ \simgt 10^{-7} {\rm M}_{\odot}$~yr$^{-1}$. 
It was shown that steady nuclear 
burning on the white dwarf surface could produce the extremely soft X-ray 
emission at the required luminosities. However, 
such high accretion rates require a donor star more massive than the white 
dwarf, in order to sustain thermally unstable mass transfer. 

We have therefore undertaken a programme of optical spectroscopy and 
photometry, 
to test the predictions of this model. In particular, we focus on the LMC 
group, since the accurately known distance of these high luminosity SSSs is a 
vital key in understanding their nature. We present here the results of 
long-term 
optical photometry of the transient LMC source, RX~J0513-69, which was 
discovered in the ROSAT All Sky Survey (Schaeidt, Hasinger \& Tr\"umper 1993).  
Remarkably, these observations were acquired as a serendipitous 
by-product of the MACHO project (Alcock et al.\ 1995a), 
owing to the location of 
RX~J0513-69 in a frequently monitored field. We are thus afforded an 
unprecedented opportunity to study the long term behaviour of this source 
which, at $V \sim 16-17$, would normally be impossible. 

\section{Observations}

The relative magnitude of RX~J0513-69 for the period 
1992 August~22 -- 1995 November~27 is presented in Fig.~1.    
The observations were made using the 
$1.27$-m telescope at Mount Stromlo Observatory, Australia. 
A dichroic beamsplitter and filters provide simultaneous CCD photometry 
in two passbands, a `red' band ($\sim 6300-7600$~\AA) 
and a `blue' band ($\sim 4500-6300$~\AA). However, we show here only the 
observations taken through the latter filter, 
which is approximately 
equivalent to the Johnson $V$ passband, 
since the red light curve is 
not appreciably different. 
The vertical dotted 
lines indicate our working definition of the `high' and `low' state 
sections, as used in the power spectrum analysis. 

The images were reduced with the standard 
MACHO photometry code {\sc SoDoPHOT}, based on 
point-spread function fitting and 
differential photometry 
relative to bright neighbouring stars. 
One-sigma error bars are shown, 
which are usually smaller than the data points ($\approx 0.02$ mag);  
for some observations with very poor seeing or transparency, 
the errors rise to $\sim 0.1$ mag. 
Further details of the instrumental set-up and data processing 
may be found in Alcock et al.\ 1995b, Marshall et al.\ 1994 and 
Stubbs et al.\ 1993. 

The absolute calibration of the MACHO fields 
and transformation to standard passbands is not yet complete, 
thus the measurements are plotted differentially relative to the 
observed median. 
A preliminary estimate indicates that
$\Delta m = 0$ corresponds to 
$V\approx 16.3$ and $V-R\approx 0.08$, with 
systematic uncertainties around 0.2 and 0.1~mag respectively.  
The final calibration will greatly reduce these uncertainties, but 
they do not affect the differential measurements presented here.

The system exhibits pronounced optical variability, with the most dramatic 
changes 
occurring on timescales of $\sim 100-200$~d. The brightness 
typically drops by $\sim0.8-1.0$~mag in only 
$\sim10$~d, after which the light curve maintains a low level for 
$\sim 30$~d. It should be noted, however, that following the initial drop, 
the system usually 
brightens by $0.3-0.4$~mag in the first $\sim8$~d, and 
maintains this plateau level before making the rapid upward transition back 
to the high state. This pattern of variability is less obvious in the low 
state at day number $\sim$ 1540, in which the system appears to undergo a 
minor outburst, rather than exhibiting the step-like behaviour. 

The first and final low states of Fig.~1 exhibit a somewhat more gradual 
decrease in magnitude than the others, taking around a week longer to 
drop to the faintest level. The overall drop is also relatively 
smaller by $\geq 0.1$~mag. The magnitude at which the system starts to enter 
a low state appears essentially constant ($\Delta m =0$), except perhaps 
for the faint episode at day number $\sim$ 900, which begins at 
$\Delta m =-0.2$. 
During the high states, the brightness appears to show a roughly linear fading, 
amounting to $\sim 0.2$~mag in total, independent of the interval of time 
between the low states. 
We note that it is probable, 
both from the recurrence times of the low states 
and from the decline of the light curve before day $\sim$ 1320, that a low 
state was missed in the period $\sim$ 1320$-$1370, during which there is a 
gap in the monitoring. Indeed, spectroscopic observations obtained in 
December~1992 by Reinsch (in preparation) strongly suggest that RX~J0513-69 
was in an optical low state during this time.

\section{Period Analysis}

We analysed the high state sections, as defined by the vertical dotted lines 
in Fig.~1, for any periodic behaviour. 
Due to the fact that the intrinsic variability of RX~J0513-69 of up to 
$\sim 0.3$~mag 
in the `flat' regions of the light curve can mask orbital modulations, 
we detrended each high state section separately, by subtracting a linear fit, 
before performing a power spectrum 
analysis on the combined dataset. 
The resulting Lomb-Scargle (Lomb 1976; Scargle 1982) periodogram is 
shown in Fig.~2, for a frequency space of $0.1-10$~cycles~d$^{-1}$, with a 
resolution of 0.001~cycles~d$^{-1}$. 
We see a dominant peak at $P=0.76278\pm0.00005$~d, with lesser power at 
3.22~d and 1.45~d. However, the former of these secondary peaks is almost 
certainly a one-day alias, having a frequency of 1 cycle~d$^{-1}$ less than 
that of the
0.76278~d period. Furthermore, we checked the significance of the 
peaks by analysing the power spectra of randomly 
generated datasets, using the sampling intervals of the real data. 
Our Monte Carlo simulations reveal that the 1.45~d peak is significant 
at only the $\sim 1.6 \sigma$ level. However, the power at 
$P=0.76278\pm0.00005$~d corresponds to a $5\sigma$ detection, 
leading us to 
present this as the true orbital period.  
Independent spectroscopic studies (Crampton et al.\ 1996) 
are consistent with this result, 
having revealed a `best' period of $P \approx 0.76$~d. 
 
The high state data were folded on $P=0.76278$~d to examine the form of the 
orbital modulation. We find that the data are well fitted by a 
sinusoid of semi-amplitude 0.0213$\pm$0.0009 mag, suggesting a low 
inclination. This phase-averaged, folded light curve is shown in Fig.~3. 
We derive 
an ephemeris of T$_{\rm o}={\rm JD}\,2448857.832(5) + 0.76278(5)E$, 
where T$_{\rm o}$ is the 
time of maximum optical brightness, and $E$ is an integer. 

\section{Discussion}

Recently, Crampton et al.\ (1996) reported variations of 
only $\sim0.3$~mag, 
noting that the optical counterpart, identified as HV~5682, has historically 
shown variations of up to $\sim1$~mag. Placing these observations in the 
context of the MACHO light curve, it is clear that both findings are 
consistent with the data presented here. In the more usual optical high 
states, variations of up to 
$\sim0.3$~mag commonly occur on timescales of days (see Fig.~1). 
Clearly, the more 
dramatic variation of $\sim 1$~mag can be identified with transitions to the 
rather more infrequent low states, detection of which requires extended 
monitoring of the type reported here. 

The optical luminosity in this system is expected to be dominated by the 
EUV/soft X-ray heated 
accretion disc, which has an absolute visual magnitude 
$M_{V} \approx -2$ in the high state, at the high extreme of 
LMXBs in general (van Paradijs \& McClintock 1995). Indeed, it is remarkable 
that {\em all} the LMC LMXBs are optically so luminous. 
This may be compared to typical values 
of $M_{V} = +4$ to $+7$ for CVs (van Paradijs 1983). 
Even if we apply the Warner (1987) empirical relation:
\begin{equation}
M^{\rm{\scriptsize{disk}}}_{{\small{v}}} = 5.74 - 
0.259~{\rm P}_{\small{orb}}({\rm hr}) \hspace{1cm} ({\rm P}_{\small{orb}} 
\simlt 15~{\rm hr})
\end{equation}
for the brightness of 
dwarf novae discs at maximum, using $P_{\small{orb}} \approx 18$~hr, 
we obtain $M_{V} = +1.1$, substantially fainter than observed 
for RX~J0513-69. This suggests that there is an additional source of 
optical luminosity in the system, consistent with the van den Heuvel 
et al.\ (1992) scenario of a white dwarf undergoing surface nuclear 
burning. 

Using our orbital period of $P=0.76278$~d, we may calculate the 
mean density, $\overline{\rho}$, of the companion star under the 
assumption that it fills its Roche lobe. We combine Kepler's 
third law and 
the Eggleton (1983) relation for a Roche-lobe filling star: 

\begin{equation}
\frac{R_{L_2}}{a}=\frac{0.49q^{-2/3}}{0.6q^{-2/3} + 
\mbox{ln}(1+q^{-1/3})}, 
\end{equation}
where $R_{L_2}$ is the radius of a sphere with the same volume as the 
secondary Roche lobe, $a$ is the binary separation, and $q$ is the binary 
mass ratio ($\equiv {\rm M}_{\scriptsize{compact}}/
{\rm M}_{\scriptsize{secondary}}$) to obtain:

\begin{equation}
\overline{\rho} = \frac{0.161}{{\rm P}^2(1+q)} \left 
(0.6 + q^{2/3} \ln(1+q^{-1/3}) \right )^3 {\rm g}~{\rm cm}^{-3}, 
\end{equation} 
where $P$ is in days. 
For values of $q \simlt 1$, as 
required by the van den Heuvel et al.\ 1992 model, the implied mean 
density is $\sim 0.2-0.3$~g\,cm$^{-3}$. This is consistent with that of a 
$\sim 2.5-3.0~{\rm M}_{\odot}$ main sequence star (of spectral type $\sim$ A0). 
The light from such a star would still 
be dominated by the accretion disk/compact object luminosity, since for an LMC 
distance modulus of 18.5 (e.\ g.\ Panagia et al.\ 1991) its apparent 
magnitude would be $\sim 19$, significantly fainter than the 
observed brightness of RX~J0513-69 of $V \sim 17$. 
 
\section{Summary}

In conclusion, we have used the remarkable monitoring capability available 
as a by-product of the MACHO project to observe, 
for the first time, the long term optical 
behaviour of the recurrent SSS RX~J0513-69. We see marked high and low states 
in the optical. 
The relation of these optical variations to the X-ray 
behaviour is currently under investigation (Southwell et al.\ 1996). 
Indeed, ROSAT has so far discovered three X-ray on states for this 
transient source, 
making it the first SSS to exhibit recurrent outbursts (Hasinger 1994; 
Schaeidt, in preparation).  
We derive an 
orbital period of $P=0.76278\pm0.00005$~d, consistent with independent 
spectroscopic findings. Such a binary system allows solutions for the nature 
of the secondary star that are consistent with the van den Heuvel et al.\ 
(1992) model of unstable mass transfer onto a white dwarf. 
Given the small 
photometric orbital modulation ($\sim 0.02$~mag semi-amplitude) 
and the extreme optical variability of this 
source, we strongly urge further spectroscopic observations. 

\subsection*{Acknowledgments}

We are grateful for the support given our project by the technical
staff at the Mt. Stromlo Observatory. Work performed at LLNL is 
supported by the DOE under contract W-7405-ENG. Work performed by the
Center for Particle Astrophysics personnel is supported by the NSF 
through grant AST 9120005. The work at MSSSO is supported by the Australian
Department of Industry, Science and Technology. 
KG acknowledges support from DoE OJI, Alfred P. Sloan, and Cotrell Scholar 
awards. 
CWS acknowledges the generous support of the Packard and Sloan Foundations.
WS and KAS are both supported by PPARC through an Advanced Fellowship and 
studentship respectively. 
KAS is grateful to T.\ Shahbaz for a useful discussion on the power spectrum 
analysis, and we thank the anonymous referee for comments that helped clarify 
the discussion.

\begin{figure}
\caption{The optical light curve of
RX~J0513--69 from 639 observations taken during the MACHO project. The 
relative magnitude is shown for the `blue' filter, which is approximately 
equivalent to the Johnson $V$ passband. Note the quasi-regular magnitude drops 
of $\sim 1$~mag. The day number is JD$-$2448000. The vertical dotted 
lines indicate our working definition of the `high' and `low' state 
sections, as used in the power spectrum analysis. }
\end{figure}

\begin{figure} 
\caption{The Lomb-Scargle periodogram of RX~J0513-69 MACHO time 
series data, excluding the low states. 
The strongest peak is at $0.76278\pm0.00005$~d.}
\end{figure}
 
\begin{figure}
\caption{The light curve of RX~J0513-69, folded on a period of 
0.76278~d. The data have been averaged into 30 phase bins, and 
are fitted with a sinusoid of amplitude 0.0213~mag. Two cycles 
are plotted for clarity. }
\end{figure}

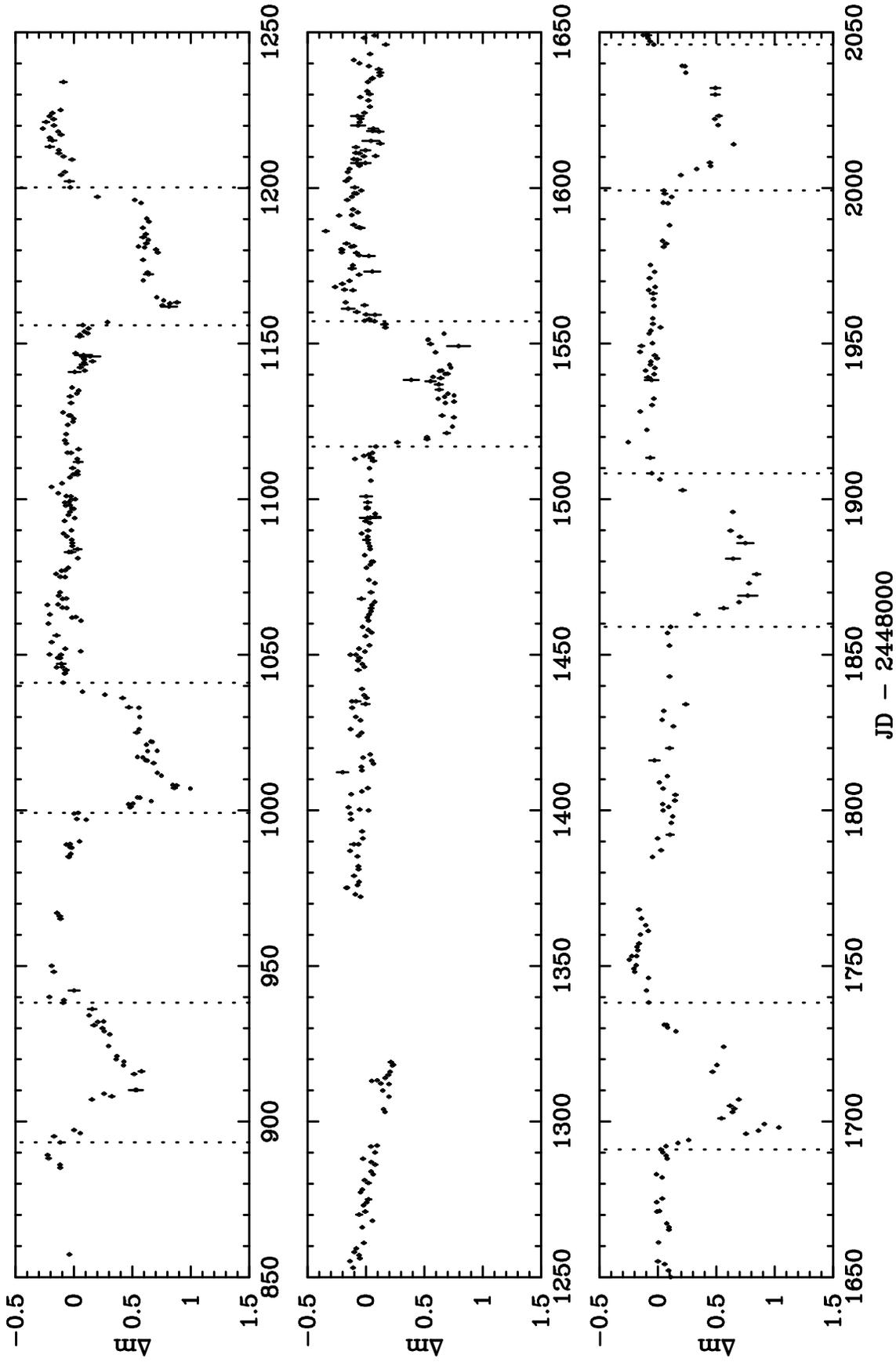
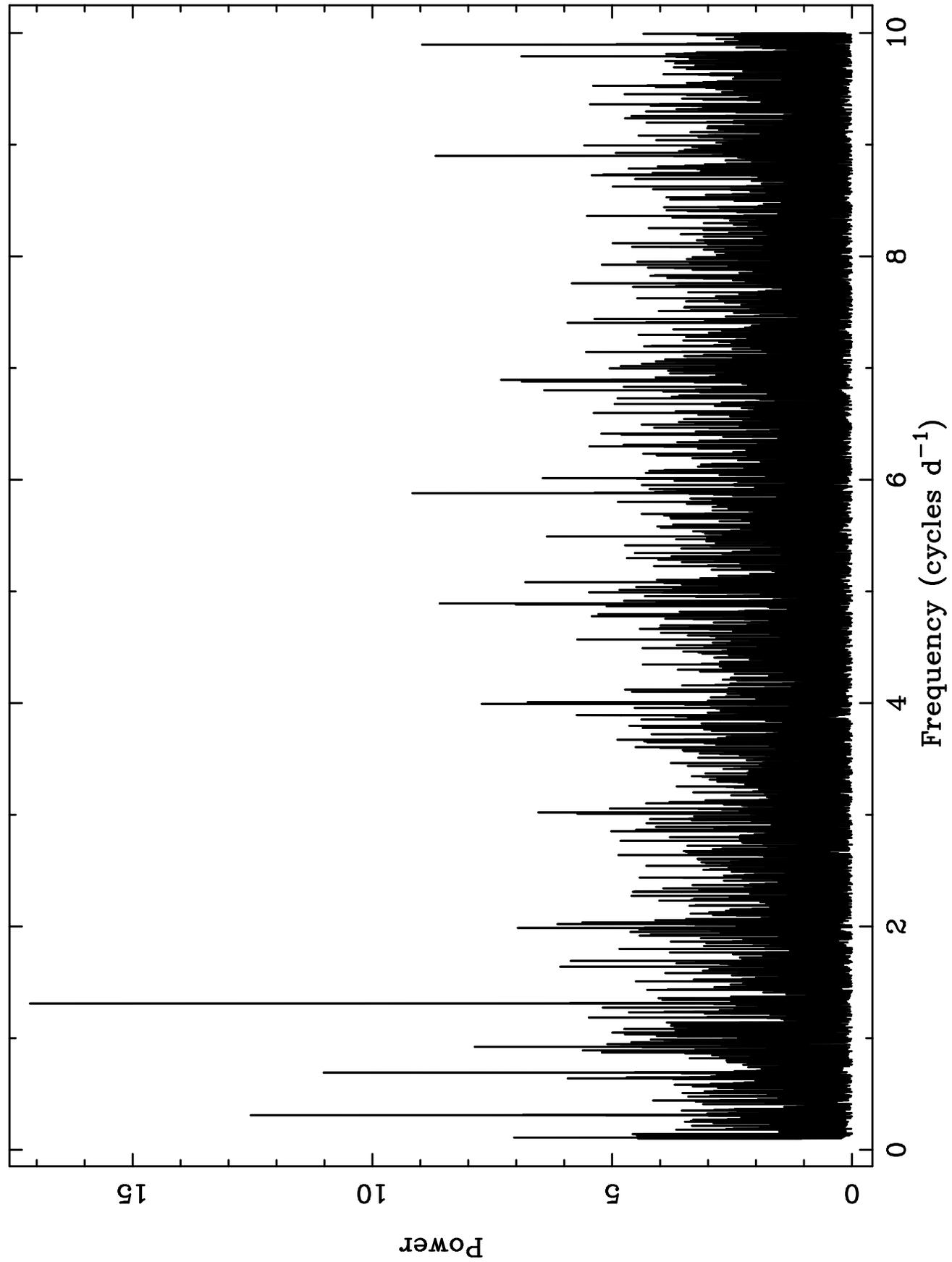
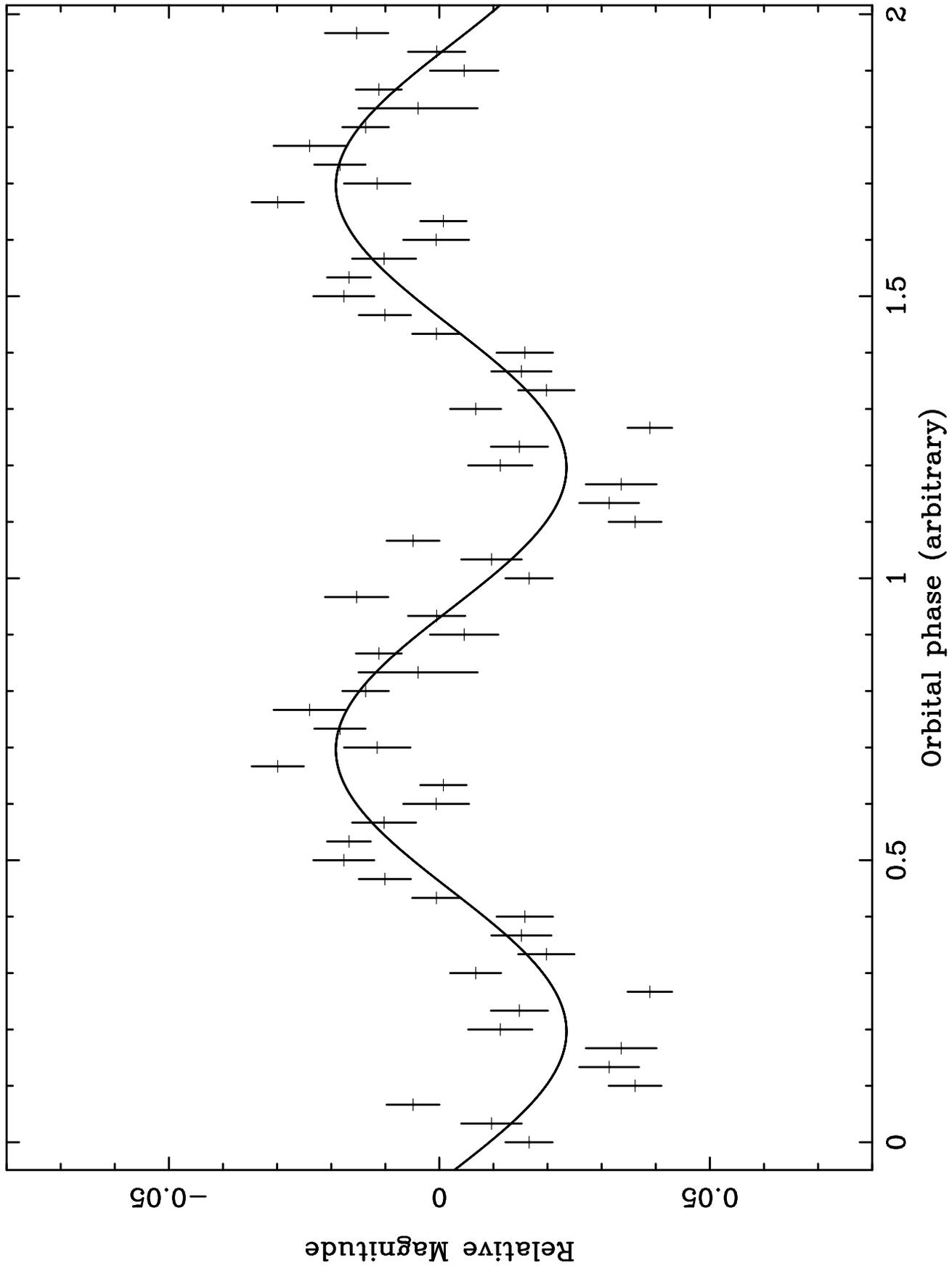
\end{document}